\documentclass[12pt,preprint]{aastex}

\usepackage{emulateapj5}
 
\newcommand{\et}{et al.}

\newcommand{\fv}{F_{var}}
\newcommand{\fvar}{F_{var}}

\newcommand{\mbh}{M_{{\rm BH}}}

\newcommand{\xte}{{\it RXTE}}

\newcommand{\asca}{{\it ASCA}}
\newcommand{\Msun}{\hbox{$\rm\thinspace M_{\odot}$}}
\slugcomment{}
\shorttitle{X-ray timing of NGC~4258}
\shortauthors{Markowitz \& Uttley}
                                      
\begin{document}\title{Low-Luminosity AGN as analogues of Galactic Black Holes in the low/hard state:
Evidence from X-ray timing of NGC 4258}
                                       
\author{A.~Markowitz\altaffilmark{1,2}, P.~Uttley\altaffilmark{1,2}
\altaffiltext{1}{X-ray Astrophysics Laboratory, Code 662, NASA/Goddard Space Flight Center, Greenbelt, MD 20771; agm@milkyway.gsfc.nasa.gov; pu@milkyway.gsfc.nasa.gov}
\altaffiltext{2}{N.A.S./N.R.C.\ Research Associate}}

\begin{abstract}

We present a broadband power spectral density function (PSD) measured from extensive 
\xte\ monitoring data of the low-luminosity AGN NGC 4258, which has an accurate, maser-determined
black hole mass of $(3.9\pm0.1)\times 10^{7}$~$\Msun$.
We constrain the PSD break time-scale to be greater than 
4.5 d at $>$90$\%$ confidence, which appears to rule out 
the possibility that NGC~4258 is an analogue of
black hole X-ray binaries (BHXRBs) in the high/soft state. 
In this sense, the PSD of NGC~4258 
is different to those of some
more-luminous Seyferts, which appear similar to the PSDs of 
high/soft state X-ray binaries. 
This result supports previous analogies between LLAGN and X-ray binaries in the
low/hard state based on spectral energy distributions,
indicating that the AGN/BHXRB analogy is valid across a broad range
of accretion rates.

\end{abstract}

\section{Introduction}

The aperiodic X-ray variability in Seyfert Active Galactic 
Nuclei (AGN) has been well-quantified over multiple timescales. 
Seyfert broadband fluctuation power spectral density functions (PSDs)
show characteristic breaks at temporal
frequencies corresponding to timescales of a few days or less (Uttley, M$^{\rm c}$Hardy \& 
Papadakis 2002, Markowitz \et\ 2003, M$^{\rm c}$Hardy \et\ 2004). 
Markowitz \et\ (2003) and M$^{\rm c}$Hardy \et\ (2004) have shown that the PSD break 
timescales measured so far are
consistent with scaling roughly linearly with black hole mass $\mbh$. Remarkably,
the mass--timescale relation is consistent with extrapolation to stellar-mass
black hole X-ray binaries (BHXRBs), and AGN and BHXRB broadband PSD shapes
are similar, suggesting that a similar variability process
is at work over at least 6 decades in black hole mass.

In BHXRBs, the PSD shape and characteristic break timescales are known to correlate
with the X-ray spectral state, which is thought to depend on the global 
accretion rate relative to Eddington, $L/L_{{\rm Edd}}$ (e.g. see reviews by 
McClintock \& Remillard 2003, van der Klis 2004). In general, shorter
characteristic timescales are correlated with steeper X-ray power-law components and
stronger thermal emission and reflection components, which suggests a model
where the inner edge of the accretion disk moves inwards at higher accretion rates
and hence characteristic timescales become shorter (e.g. Gilfanov, Churazov \& Revnivtsev
1999).  The difference in characteristic timescales is clear when comparing the PSDs of the
low/hard spectral state, in which PSD breaks appear at
a few Hz, with the high/soft state, where a similar steepening occurs at higher frequencies 
$>10$~Hz (e.g. Churazov, Gilfanov \& Revnivstev 2001).

Recently, M$^{\rm c}$Hardy \et\ (2004) noted that higher accretion rate Narrow Line 
Seyfert~1 (NLS1) AGN  appear to have relatively shorter PSD break timescales for 
their black hole mass than normal Seyferts (and also appear most similar
to the PSDs of BHXRBs in the high/soft state, M$^{\rm c}$Hardy et al. 2004, 2005).  Since NLS1
are thought to accrete at high rates (Pounds, Done \& Osborne 1995), 
this observation would suggest that
characteristic timescales in AGN follow a similar dependence on accretion rate to those
in BHXRBs.  However, the accretion rates of AGN with good-quality PSDs measured so
far are estimated to be a few per cent of Eddington or greater.  Since
the low/hard--high/soft state transition in BHXRBs seems to occur at $\sim2\%$
of Eddington (Maccarone 2003), it is not yet clear whether any AGN observed so far should be in
the low/hard state (with the possible exception of NGC~3783, Markowitz et al. 2003).  
Therefore, in order
to better test the analogy with BHXRBs and search for evidence of different
states in AGN, it is useful to measure the PSDs of X-ray light curves of
AGN accreting at significantly lower rates than the existing sample,
which may be analogues of BHXRBs in the low/hard state.

In this Letter, we make a first step towards measuring a high-quality PSD for
a low-accretion rate AGN, by measuring a preliminary
PSD for the LLAGN NGC~4258 ($L/L_{{\rm Edd}}$$\sim$10$^{-4}$, e.g., Lasota \et\ 1996), 
using monitoring data obtained in 1997--2000
by the {\it Rossi X-ray Timing Explorer} ({\it RXTE}).  
NGC~4258's black hole mass
of $(3.9\pm0.1)\times 10^{7}$~$\Msun$ is highly accurately measured via
VLBI studies of its water megamaser (Miyoshi \et\ 1995, Herrnstein \et\ 1999),
allowing us to make an accurate comparison with the PSD expected from both
high/soft and low/hard accretion states.

\section{Observations and Data Reduction}

We constructed a high-dynamic-range PSD for NGC~4258 by combining 
monitoring on complementary timescales (e.g., Edelson \& Nandra 1999). 
NGC~4258 was monitored regularly once every 2--3 days by \xte\ from 
1997 Dec 27 to 2000 Mar 1 (``long-term'' monitoring). 
Each visit lasted $\sim$1~ksec. There was also intensive, simultaneous \xte\ 
and \asca\ monitoring during 2000 May 15--20, when the source was observed
every orbit (``short-term'' monitoring).  For consistency we only use the
short-term \xte\ PCA data here, binned to 5760~s (i.e. single orbit) 
intervals, but note that the 2--10~keV
\asca\ short-term light curve is consistent with the \xte\ light curve. 
Data were obtained from {\it RXTE}'s proportional counter array (PCA),
using standard extraction methods and selection criteria appropriate for faint sources
(see Markowitz \et\ 2003 for further details). 
Light curves were generated over the 2--10 keV bandpass. 
All count rates quoted herein are normalized to 1 proportional counter unit (PCU). 
The light curves are plotted in Figure~1. 
One can see that NGC~4258 displays minimal variability on short timescales; 
other observations (e.g., Pietsch \& Read 2002) also 
show minimal variability on $\lesssim$day timescales. In 
contrast, there is strong variability on long timescales. The fractional 
variability amplitude $\fv$, as defined in e.g., Vaughan \et\ (2003), is 4.2 $\pm$ 0.8 $\%$
and 27.9 $\pm$ 0.2 $\%$ for the short- and long-term light curves,
respectively\footnote{The
errors on $\fv$ account for observational noise only, not intrinsic stochastic variability.}.

The 2--10 keV X-ray emission is
dominated by the direct nuclear emission in this source (e.g., Fiore \et\ 2001).
Variability due solely to variations in the intrinsic column density is negligible,
as the column density is not strongly variable, even on
timescales of years (Risaliti, Elvis \& Nicastro 2002). Furthermore,
a plot of the binned 2--4 keV flux versus the 7--10 keV flux (which, for brevity,
we do not show here)
shows a continuous, virtually linear distribution of points, suggesting 
a lack of strong absorption events during the monitoring.

\section{The PSD of NGC~4258}

\subsection{Constraints on PSD break timescale}

We used the {\sc psresp} Monte Carlo method of Uttley \et\ 2002 (and see Markowitz 
\et\ 2003) to constrain the shape of the PSD.  Monte Carlo simulations are
required to take account of aliasing effects due to sparse and/or irregular sampling,
which distorts the PSD shape.  Furthermore, simulations
are essential to properly determine confidence
limits and goodness of fit of models while using the full dynamic range of the PSD,
since the lowest frequencies in the PSDs of the long- and short-term light curves are
not well-sampled enough to allow standard, Gaussian errors to be assigned to the PSD.
The method is fully described in Uttley \et\ (2002), Markowitz \et\ (2003) and see 
also M$^{\rm c}$Hardy (2004)\footnote{We note here that to improve S/N, the long-term
light curve was binned up in 2-week intervals.  Model fits were carried 
out using grids of
PSD model parameter values, and at each point in the grid
400 simulated light curves were generated for each light curve (long 
and short-term), resampled to match the sampling of the observed light curves
(and rebinned to 2-week bins in the case of the long-term simulated data) 
and 4000 combinations of the resulting simulated PSDs were used
to estimate the goodness of fit for that set of model parameters.}.
The long- and short-term PSDs spanned the temporal frequency ranges 
$2\times 10^{-8}$--$3\times 10^{-7}$~Hz and
$4\times 10^{-6}$--$9\times 10^{-5}$~Hz respectively.
The power due to Poisson noise over the 2--10~keV bandpass was 
860 Hz$^{-1}$ and 8.3 Hz$^{-1}$ for the 
long- and short-term PSDs, respectively. 

To constrain the location of any PSD break, we employed a broken power law
model of the form $ P(f) = A(f/f_c)^{-1},  f \le f_c$, and
$ P(f) = A(f/f_c)^{- \beta},f > f_c  $, where the normalization $A$ is the PSD
amplitude at the break frequency $f_{c}$,
$\beta$ is the high frequency power law slope, with the constraint $\beta>1 $. 
This simple model provides a good description of the PSDs of Seyfert 
galaxies measured so far, with PSD breaks detected in nine of those AGN.
The range of $\beta$ tested was 1.0--2.3 in increments of 0.1.
Break frequencies were tested from $10^{-8}$--$10^{-5}$~Hz, in multiplicative
steps of 1.5.
The best-fitting model is one with a break at $2.25\times 10^{-8}$~Hz, although 
an unbroken
power-law is also formally acceptable (i.e. the break frequency could lie
out of the observed range).  The best-fitting
$\beta = 2.3$, $A=(3.3\pm^{1.3}_{0.9})\times 10^{6}$~Hz$^{-1}$, and
the ``rejection
probability" (corresponding to the fraction of simulated PSD sets which are a better
fit to the assumed model than the real data) is 0.43, i.e. the model is
formally acceptable.
Contour plots showing the rejection probabilities for a given
$\beta$ and $f_c$ are shown in Figure 2. On the contour plot we also show
the expected break frequencies assuming linear scaling with mass from typical
values of break timescale in the low/hard and high/soft states of
the BHXRB Cyg~X-1 (assuming a 10~$\Msun$ black hole in Cyg~X-1, e.g.,
Herrero \et\ 1995,
and the well-determined maser mass of $(3.9\pm0.1) \times10^{7}$~$\Msun$ in
NGC~4258).  The high/soft state break frequency (corresponding to 4.5 d)
is ruled out at 
$>$90\% confidence, while the low/hard state break ($\sim$45 d)
is acceptable at this level of confidence.

Fiore \et\ (2001) report a 1998 December {\it BeppoSAX} 3-10~keV
light curve of NGC~4258 which shows
almost a factor 2 change in flux in one day.  This behavior is atypical of the source,
which shows much smaller variations on timescales of a few days in both the {\it RXTE} long-term
monitoring and the much more intensive monitoring.  We investigated whether this relatively large
variation in the {\it BeppoSAX} observation was consistent with the PSD derived from {\it RXTE}
monitoring, by including the {\it BeppoSAX} light curve (obtained directly from the
public archive) in our fits.   We find that the rejection probability for the broken power-law PSD is 
increased to 0.77, but the overall confidence contours for $f_c$ and $\beta$
are only made marginally wider.  We conclude that the {\it BeppoSAX} light curve is consistent
with the {\it RXTE} {\it PSD} and simply represents a statistical outlier in the stochastic
variability process.

\subsection{Comparison with other AGN}

The PSD of the LLAGN NGC~4258 does indeed appear to be better explained by
scaling from a low/hard state PSD than from a high/soft state PSD,
{\it unlike} the PSDs of some of the Seyfert galaxies measured so far
which rule out a low/hard state interpretation
and are consistent with a high/soft state interpretation
(e.g. NGC~4051, MCG-6-30-15, M$^{\rm c}$Hardy \et\ 2004, 2005; NGC~3227, Uttley \&
M$^{\rm c}$Hardy 2005).  To demonstrate this difference in the PSDs,
we show in Figure 3
a relatively model-independent comparison between the PSD of NGC~4258 and that of
a normal Seyfert with nearly identical black hole mass, NGC~3516 
($\mbh$ = 4.3$\pm$1.5~$\times$~10$^7$~$\Msun$, Peterson \et\ 2004),
together with the PSDs
of Cyg~X-1 in low/hard and high/soft states for further comparison. 
NGC~3516's accretion rate is likely $\sim$5$\%$ of Eddington, given its 2--10 keV luminosity 
of 10$^{43}$ erg s$^{-1}$ and assuming the X-ray-to-bolometric luminosity scaling of
Padovani \& Rafanelli (1988). Within the errors,
the PSD break timescale of NGC~3516 is more consistent with linearly scaling from the 
high-frequency break of Cyg X-1 in the high/soft state, though scaling from the
low/hard state cannot be ruled out. 
From the figure, although
its amplitude at low frequencies is similar to that of NGC~3516,
NGC~4258's PSD is consistent with being shifted downward in temporal 
frequency by a factor of $\sim$100 relative to NGC~3516's PSD.
Given the similarity in black hole mass between the two objects and the overlapping
temporal-frequency range of the PSDs, the
fact that a PSD break was unambiguously detected in the PSD of NGC~3516
but not in that of NGC~4258 reinforces this notion.
Furthermore, the fact that NGC~4258 shows similar levels of long-timescale variability
to normal Seyferts
of comparable mass ($\fvar$ $\sim$ 30$\%$, Markowitz \& Edelson 2004)
rules out the possibility that the normalization of NGC~4258's PSD 
is significantly lower compared to that for normal Seyferts.
The PSDs are thus consistent with the notion that for a given mass, a PSD 
shifted towards relatively
lower temporal frequencies is associated with a lower accretion rate.
NGC~4258, and by extension, other LLAGN, may thus be a low accretion rate 
version of normal Seyferts, as far as   
X-ray continuum variability is concerned.

\section{Discussion and Conclusions}

We have shown that the PSD of NGC~4258 is consistent with having the same
shape and normalization
as the PSDs of more-luminous Seyfert galaxies, but with a break on longer timescales
than expected by scaling from the high/soft state of Cyg~X-1, unlike a Seyfert galaxy
with similar mass, NGC~3516.  From a purely phenomenological point of view,
this result may help to explain the differences in short-term ($<$day) variability
amplitude between LLAGN and Seyfert galaxies reported by Ptak \et\ (1998), who
showed that LLAGN do not follow the well-known anticorrelation between
X-ray luminosity and short-term
variability amplitude which is shown by normal Seyferts (e.g., Nandra \et\ 1997), but
instead populate the low-luminosity, low-variability part of the variance-luminosity 
diagram.  NGC~4258 also shows a very low variability amplitude on short timescales,
and our PSD suggests that this is due to a long break timescale in this AGN,
rather than a low overall normalization of the PSD.  The same might also
be true of other LLAGN.  

The comparison with break-timescale expected from mass-scaling of the timescales 
from Cyg~X-1, and the comparison with the PSD of NGC~3516, both suggest
that the break timescale in NGC~4258 is intrinsically longer (given its mass)
than in more luminous AGN.  This difference may reflect the
low accretion rate in NGC~4258.
NGC~4258 appears to be consistent with scaling
from the low/hard state PSD of Cyg~X-1, although it is possible that the
break timescale is even longer than that scaling would suggest.  An analogy
with other BHXRBs which display a larger range of $L/L_{{\rm Edd}}$
than Cyg X-1 supports this possibility.  For example
in the outburst decays of transient BHXRB candidates,
after transition from the high/soft spectral state to the low/hard spectral
state, there is evidence that PSD timescales, particularly QPOs,
continue to migrate towards
lower frequencies as the luminosity and accretion rate of the source
decrease and the source approaches quiescence (van der Klis 2004, Kalemci \et\ 2004, 
Rodriguez \et\ 2004, Nowak, Wilms \& Dove 2002).

Despite the consistency with BHXRB behavior, it is still not clear whether the
dependency of the mass--timescale relation on AGN accretion rate is
a continuous scaling, or if it exists in the form of a high vs.\ low accretion rate
dichotomy akin to that seen in BHXRBs. 
However, there are differences in the observed energy-spectral features of LLAGN
and normal Seyferts which are suggestive of differing accretion modes in the two
classes of objects.
Specifically, LLAGN are more radio loud compared to
normal Seyferts, and the so-called optical/UV ``big blue bump'' in normal Seyferts, 
inferred to be thermal accretion disk emission, is absent in LLAGN (Ho 1999).
Similar discrepancies arise when comparing the spectral energy distributions of
low/hard and high/soft-state BHXRBs (though in this case, due to the much smaller,
hotter disks of BHXRBs, the thermal emission appears in the X-ray band).
Nagar \et\ (2005) have suggested that both LLAGN and low/hard state BHXRBs
are low-efficiency accreting sources, as both classes of objects are often associated 
with strong radio jet emission and low levels of thermal disk emission.
Higher-efficiency accreting sources, namely normal Seyferts and high/soft-state BHXRBs,
meanwhile, usually lack strong radio emission but have a strong disk signature, with
the supermassive systems also having gas-rich nuclei.
Additionally, a fundamental plane between $\mbh$ and X-ray and radio luminosities is seen
in both stellar-mass and supermassive systems (Merloni, Heinz \& di Matteo 2003;
Falcke, K\"{o}rding \& Markoff 2004). However, this relation breaks down for
high-state objects; radio power is reduced e.g., for high/soft state BHXRBs and
normal Seyferts (Gallo, Fender \& Pooley 2003; Maccarone, Gallo \& Fender 2003),
further supporting the notion of a similar dichotomy of states for both AGN and BHXRBs.
Finally, based on the distributions of $\mbh$ and bolometric luminosities,
Jester (2005) has suggested that in AGN, there is tentative evidence for the existence
of a critical accretion rate, 
$L/L_{{\rm Edd}}$ $\sim$ 0.01, which separates high- and low-efficiency 
accretion modes. The number of low-efficiency accreting sources in that 
sample was small, but all were LLAGN.

%%%%theoretical models:
For LLAGN and low/hard-state BHXRBs, the
longer variability timescales for a given mass, reduced emission from the optically thick, radiatively-efficient
disk, and relative weakness of
Fe K$\alpha$ emission (e.g., Pietsch \& Read 2002; Gilfanov, Churazov \& Revnivtsev
1999) are all consistent
with the suggestion that the inner accretion
disk is truncated, possibly existing at smaller radii as an optically thin, radiatively inefficient,
advection-dominated
accretion flow (ADAF; Narayan \& Yi 1995). ADAF models have frequently been 
invoked to explain the spectral energy distributions in NGC~4258 and other LLAGN
(e.g., Yuan \et\ 2002). As
one example, one popular geometrical model for the accretion configuration is
the so-called ``sphere + disk'' model of Dove \et\ (1997) and Esin, McClintock \& 
Narayan (1997). This model features a geometrically-thick, sometimes spherical, ADAF flow,
at the innermost radii, surrounded by a radiatively efficient, geometrically thin
disk. The transition radius between the ADAF and thin disk appears at
relatively larger radii for lower accretion rates.
Also, considerations of how an ADAF/coronal accretion flow can
form from evaporation of the optically thick disk have
suggested that the inner disk may evaporate completely into an optically-thin
flow if the accretion rate transitions from above to below a 
critical accretion rate (R\'{o}\.{z}a\'{n}ska \& Czerny 2001; 
Meyer-Hofmeister \& Meyer 2003 and references therein).
Further progress in this area can be made by firmly establishing whether such a
critical accretion rate exists in AGN and if it is indeed responsible for
the dominance of high- or low-efficiency accretion modes
as a function of accretion rate. Measuring PSD breaks for a larger number of
AGN, including LLAGN, is also necessary to critically test
the dependence of the mass--timescale relation on accretion rate. Clarification
of these issues in AGN as well as in BHXRBs will help to unify the behavior of
stellar-mass and supermassive accreting black holes.

%%%%%%%%%%%%%%%%%%%%%%%%%%%%%%%%%%%%%%%%%%%%

\acknowledgments The authors acknowledge useful discussions with J.\ Cannizzo.  %%%Tod, too?
This work has made use of data obtained through the High Energy
Astrophysics Science Archive Research Center Online Service, provided by
the NASA Goddard Space Flight Center, and the NASA$/$IPAC Extragalactic Database which is
operated by the Jet Propulsion Laboratory, California Institute of
Technology, under contract with the National Aeronautics and Space
Administration.

%%%%%%%%%%%%%%%%%%%%%%%%%%%%%%%%%%%%%%%%%%%%%%%% TABLES %%%%%%%%%%%%%%%

%%%%%%%%%%%%%%%%%%%%%%%%%%%%%%  FIGURES %%%%%%%%%%%%%%%%%%%%%%%%%%

\clearpage

\begin{figure}[!ht]
\epsscale{0.60}
\plotone{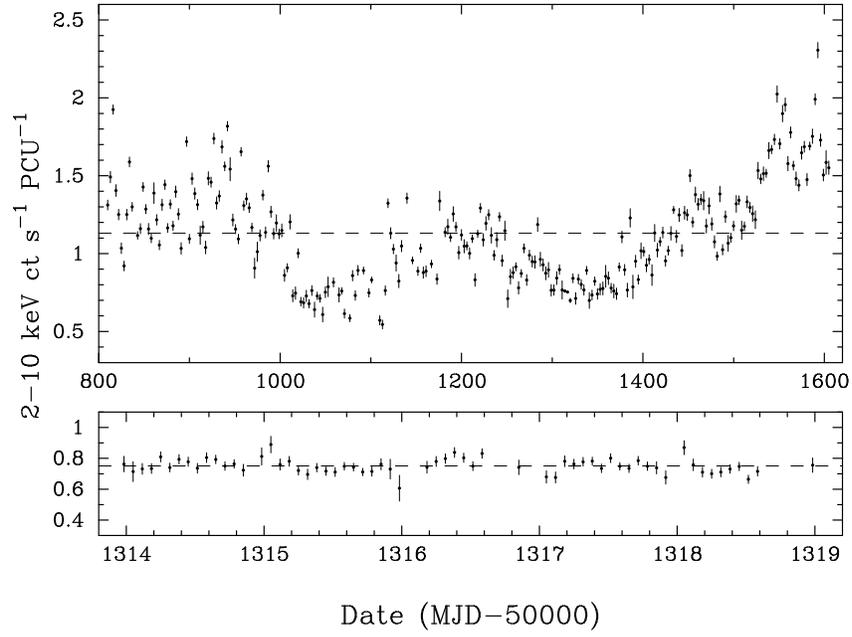}
\caption{2--10 keV light curves for the long-term (top) and short-term (bottom)
monitoring.}
\end{figure}

%\clearpage

\begin{figure}[!hb]      \vspace{-3cm}
\epsscale{0.55}
\plotone{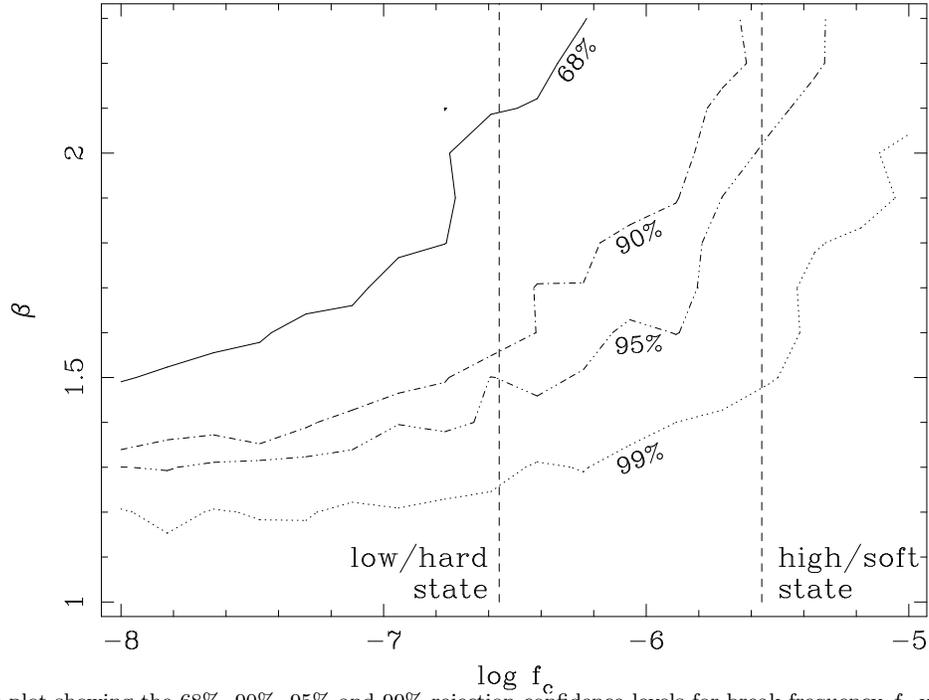}
\caption{Contour plot showing the 68$\%$, 90$\%$, 95$\%$ and 99$\%$
rejection confidence levels for break frequency $f_c$ versus high-frequency PSD slope $\beta$.
The vertical dashed lines show the break frequency predictions from linearly scaling
from the low/hard and high/soft states of Cyg X-1.}
\end{figure}

\clearpage

\begin{figure}[hb]
\epsscale{0.60}
\plotone{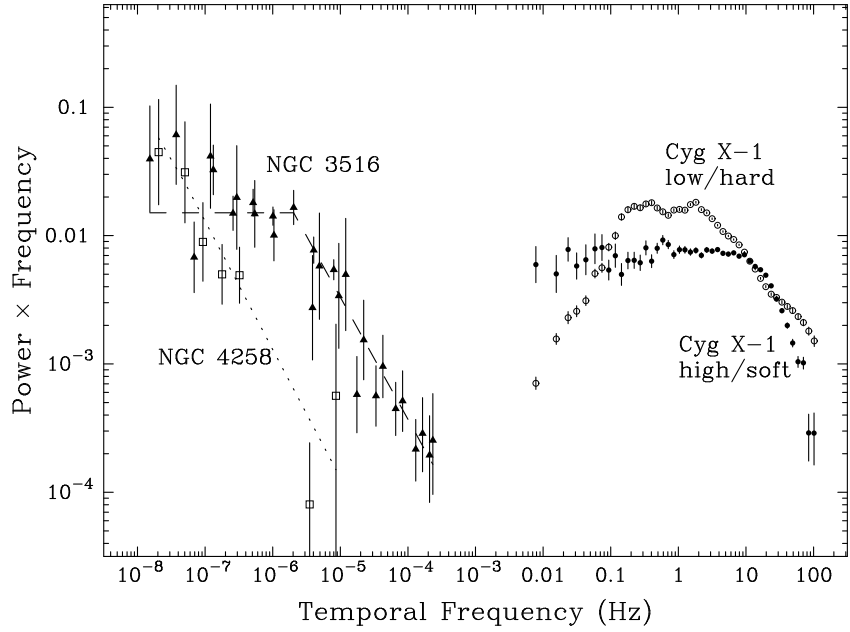}
\caption{PSDs of NGC~4258 and NGC~3516. As discussed in the text, the two AGN have similar black
hole masses, but the PSD of the lower accretion rate NGC 4258 is consistent with being scaled
to lower temporal frequencies. 
For comparison, we also plot the PSDs of Cyg X-1 in the low/hard and high/soft states.
The PSDs are plotted in Power~$\times$~Frequency to allow an easy comparison between sources.
The five highest-frequency short-term PSD points for NGC~4258 
are very poorly constrained because they are close to the Poisson noise level, and are therefore not plotted.}
\end{figure}

\end{document}